\documentclass[
pra,
 amsmath,
 amssymb,
reprint,%
superscriptaddress
]{revtex4-1}

\usepackage{amsfonts}
\usepackage{amssymb}
\usepackage{graphicx}
\usepackage{physics}
\usepackage{amsmath}
\usepackage{bm}
\usepackage{placeins} 
\usepackage{hyperref}
\DeclareMathAlphabet{\mathcal}{OMS}{cmsy}{m}{n}
\usepackage{svg}
\usepackage[english]{babel}
\usepackage{color}
\usepackage[version=3]{mhchem} 
\usepackage[retain-explicit-plus=true]{siunitx}   
\usepackage{booktabs} 
\usepackage{dcolumn}
\usepackage[utf8]{inputenc}
\usepackage[T1]{fontenc}
\usepackage{mathptmx}
\usepackage{etoolbox}
\usepackage{ulem}  
\renewcommand{\ket}[1]{\bigl | #1 \bigr \rangle}
\renewcommand{\bra}[1]{\bigl\langle #1 \bigr|}

\newcommand{\Lket}[1]{| #1  \rangle \!\rangle}
\newcommand{\Lbra}[1]{\langle \!\langle#1 |}
\newcommand{\Lbraket}[2]{\langle\! \langle#1 | #2 \rangle\!\rangle}

\newcommand{\Fket}[1]{|\!\!\!| #1  \}\!\!\!\} }
\newcommand{\Fbra}[1]{\{\!\!\!\{ #1 |\!\!\!|}
\newcommand{\Fbraket}[2]{\{\!\!\!\{ #1 |\!\!\!| #2 \}\!\!\!\} }
\newcommand{\Tket}[1]{| #1  \}}
\newcommand{\Tbra}[1]{\{ #1 |}

\newcommand{\TOp}[1]{\mathsf{#1}}

\newcommand{\diff}{\mathrm{d}}
\newcommand{\eul}{\mathrm{e}}
\newcommand{\ii}{\mathrm{i}}

\makeatletter
\def\@email#1#2{%
 \endgroup
 \patchcmd{\titleblock@produce}
  {\frontmatter@RRAPformat}
  {\frontmatter@RRAPformat{\produce@RRAP{*#1\href{mailto:#2}{#2}}}\frontmatter@RRAPformat}
  {}{}
}%
\makeatother
\begin{document}

\title{Uniform process tensor approach for the calculation of multi-time
correlation functions of non-Markovian open systems}

\author{Matteo Garbellini}
\affiliation{
Institute of Theoretical Physics, TUD Dresden University of Technology, 01062 Dresden, Germany
}
\email{matteo.garbellini@tu-dresden.de}
\author{Konrad Mickiewicz }
\affiliation{
Institute of Theoretical Physics, TUD Dresden University of Technology, 01062 Dresden, Germany
}
\author{Valentin Link}
\affiliation{Institut für Physik und Astronomie, Technische Universität Berlin, D-10623, Berlin, Germany
}
\author{Alexander Eisfeld}
\affiliation{
Institute of Theoretical Physics, TUD Dresden University of Technology, 01062 Dresden, Germany
}
\affiliation{Max Planck Institute for the Physics of Complex Systems, N\"{o}thnitzer Str.\,38, 01187 Dresden, Germany}

\author{Walter T. Strunz}
\affiliation{
Institute of Theoretical Physics, TUD Dresden University of Technology, 01062 Dresden, Germany
}

\begin{abstract}\noindent
The process tensor framework to open quantum systems provides the most general description of multi-time correlations in non-Markovian quantum dynamics. A compressed representation of a process tensor in terms of matrix product operators (MPO) can be used for numerically exact calculations of multi-time correlation functions in systems strongly coupled to a non-Markovian reservoir. We show here that the numerical scaling for computing multi-dimensional spectra can be significantly improved using a time-translation invariant MPO representation of the process tensor obtained from the uniform time-evolving matrix product operator (uniTEMPO) method. In particular, this approach provides a spectral representation of the non-Markovian dynamics that gives direct access to correlation functions in Fourier-space, avoiding explicit real-time evolution. We calculate linear and 2D electronic spectra for an example system and discuss the performance and numerical scaling of our simulations.

\end{abstract}
\maketitle

\section{Introduction\label{sec:introduction}}

Multi-time correlation functions  are of fundamental importance to probe properties of quantum systems.
They appear in multi-time measurement statistics, spectroscopy, or in the response of a system to several (external) perturbations that occur at different times.
They are used for example in the investigation of  photon bunching and anti-bunching \cite{Brown1954, Kimble1977}, electron spin resonance, nuclear magnetic resonance \cite{Ernst1990},
neutron scattering \cite{Lovesey1984}, Raman spectroscopy  \cite{cao_10.1063/1.1445746}, electron transport in quantum dot systems, exciton dynamics in nanosystems \cite{Amgar2019},  coupled cavity polariton systems \cite{Zhang2016} or  assemblies of Rydberg atoms \cite{Mukherjee_2020}.
An important experimental technique, whose theoretical modeling is commonly based on multi-time correlation functions, is femtosecond electronic spectroscopy  of molecular systems  \cite{Mu95__,Cho_doi101021cr078377b,MaKu11__, Kuhn2020, Krich2025}.
In multi-dimensional spectroscopy, the external perturbations are laser pulses applied at different times.
Typically, one transforms two time intervals between the pulses to frequency space, resulting in two-dimensional spectra which depend on the remaining time intervals.
In the following we refer to these spectra as 2D-spectra. Multi-dimensional spectroscopy has been used, for example, extensively to investigate energy transfer in biological light harvesting systems  \cite{BrMaSt04_4221_,Cho_doi101021cr078377b}.

In all the examples above, typically the relevant degrees of freedom which are probed are coupled to a large environment.
For example in the case of multi-dimensional electronic spectroscopy of molecules, this environment usually consists of certain molecular vibrations and the liquid or protein surroundings.
Since one is not interested in these degrees of freedom, open quantum system methods thus constitute an adequate framework for this analysis.
In particular, the process tensor is employed for multi-time properties \cite{Pollock2018Jan}.

If the environment is memory-less (Markovian), then the response function can be obtained via Lindblad-type quantum master equations and quantum regression theory.
However, typically the environments are non-Markovian, i.e.\ they show memory effects.
Then one has to resort to non-Markovian methods, which greatly complicates the calculations \cite{makriTensorPropagatorIterative1995, WangThoss2003, Hughes2009Jul, cerrilloNonMarkovianDynamicalMaps2014, suessHierarchyStochasticPure2014, tanimuraNumericallyExactApproach2020}.

For the case of  multi-dimensional spectroscopy such non-Markovian calculations have been performed for example using the hierarchical equations of motion (HEOM) \cite{IsFl09_234111_,Dijkstra_10.1063/1.4917025,hoshino2026heombasednumericalframeworkquantum} or the non-Markovian quantum state diffusion (NMQSD) \cite{Chen_10.1063/5.0107925,ZhEi16_4488_} or various other methods (for several examples see e.g.~Ref.~\cite{Gelin_doi:10.1021/acs.chemrev.2c00329}).

In recent years a new class of methods for simulating non-Markovian open system dynamics has been actively developed that aims at finding efficient representations of the dynamics by expressing the environment influence functional \cite{feynman63}, now commonly referred to as a process tensor \cite{Jorgensen2019Dec}, as a matrix product operator (MPO). 
Examples of this class are the PT-TEMPO (process tensor time-evolving MPO) method \cite{Pollock2018Jan,Jorgensen2019Dec} and its time-translation invariant version, the uniform TEMPO (uniTEMPO) \cite{Link_PhysRevLett.132.200403}. 
In this approach, temporal correlations in the dynamics are systematically "compressed'' by making use of tensor network theory \cite{banulsMatrixProductStates2009a,Strathearn2018Aug, leroseInfluenceMatrixApproach2021,Cygorek2021Jan, Sonner2021Dec}. For Gaussian bosonic \cite{Jorgensen2019Dec, fuxEfficientExplorationHamiltonian2021, Link_PhysRevLett.132.200403} and fermionic \cite{Thoenniss2023May,ngRealtimeEvolutionAnderson2023,Chen2024Jan, Sonner2025Oct} environments, highly efficient algorithms exist that achieve this compression, allowing for a subsequent fast evaluation of open system dynamics. 
This approach has been applied, for example, to equilibrium dynamics \cite{Nguyen2024Sep}, (optimal) control \cite{fuxEfficientExplorationHamiltonian2021, Kahlert2024Nov}, Floquet dynamics \cite{Mickiewicz2025Nov}, quantum thermodynamics \cite{Gribben2022Feb}, quantum memory \cite{dowlingCapturingLongRangeMemory2024,backerVerifyingQuantumMemory2025} and dynamical mean-field theory \cite{nayakSteadystateDynamicalMean2025}. 
It is also very well suited for computing multi-time correlation functions \cite{Gribben2022Feb,Fux2023Aug,salamonMarkovianApproachNphoton2025b}, and has recently been applied to multi-dimensional electronic spectroscopy \cite{deWitt_PhysRevResearch7013209}. 
Most algorithms for generating process tensors in matrix product operator form use a finite time window, limiting the maximum evolution time and obscuring the time-translation invariance inherent in the model. This caveat is lifted in the uniTEMPO method, which uses infinite tensor network methods to encode a dynamical semi-group structure into the MPO representation of the influence functional, reflecting the stationarity of the correlation function \cite{Link_PhysRevLett.132.200403,Sonner2025Oct}. 
We show in this work that this form is highly effective for computing multi-time correlation functions in non-Markovian open systems. In particular, corresponding spectra can be obtained directly through a spectral representation, avoiding real-time evolution and thus achieving a superior numerical scaling.

The paper is organized as follows:
In section \ref{sec:MultCorr} we provide the definition of multi-time correlation functions and specify the open quantum system model.
In section \ref{sec:UniTEMPO} we show how to calculate multi-time correlation functions using uniTEMPO, which is the most important part of the present work.
In section \ref{sec:2D-spek} we provide explicitly the correlation functions and the corresponding frequency-domain expressions used for multi-dimensional-spectroscopy.
In section \ref{sec:2D-unitempo} 2D-spectra are calculated for a simple model system using the uniTEMPO method.
Finally in section \ref{sec:conclusions} we conclude and give a brief outlook.

Throughout the work we set $\hbar=1$ and we also set the Boltzmann constant $k_\mathrm{B}=1$.

\section{Multi-time correlation functions\label{sec:MultCorr}}

\subsection{General definition\label{sec:general_def}}

The $N$-time correlation function is written as

\begin{equation}
\label{eq:correlation_general}
    R(\tau_N,\dots,\tau_1)
    =
    \Lbra{\openone}\mathcal{V}_{N+1} \mathcal{U}(\tau_N)\mathcal{V}_{N} \cdots \mathcal{V}_{2} \,\mathcal{U}(\tau_1)\mathcal{V}_{1} \Lket{\rho_0}.
\end{equation}
Here we have used the Liouville-Space notation where the density operator corresponds to a vector $|\rho\rangle\rangle$ and $\Lbraket{A}{B}=\mathrm{Tr}[A^\dagger B]$, thus $\Lbraket{\openone}{\rho}=\mathrm{Tr}[\rho]$.
The operators $\mathcal{U}(\tau_j) = e^{-\ii \mathcal{L} \tau_j}$  describe  time evolution of duration $\tau_j$.
One has the correspondence $\mathcal{U}(\tau_j)\Lket{\rho} \leftrightarrow U(\tau_j)\rho U^\dagger(\tau_j)$ with $U(\tau)=\eul^{-\ii H \tau}.$
The Liouville-space operators $\mathcal{V}_j$ describe the $j$th interaction with the system.
In a density matrix formulation the corresponding  Hilbert space operator $V_j$ could act either on the bra or on the ket side of the density matrix, which results in different $\mathcal{V}_j$, as will be explained in detail below.
A diagrammatic representation of Eq.~(\ref{eq:correlation_general}) is given in Fig.~\ref{fig:diagrams}a) and b).

Often a frequency space representation facilitates the interpretation of the correlation function. 
Such a representation is achieved by a (half-sided) Fourier transformation of one or more time intervals $\tau_j$.
We will provide explicit examples, when discussing the multi-dimensional spectroscopy below.

\subsection{Open quantum system formulation}

In many situations the system can be described by a small number of degrees of freedom which in turn are coupled to a (large) environment.
For example in the electronic spectroscopy of molecules, considered below, the system can be described by a few electronic states, which are coupled via the laser pulses, and the environment consists of molecular vibrational modes and the molecular surrounding \cite{Mu95__,MaKu11__}.
In such a situation, where the interaction operators $\mathcal{V}_j$ do not directly act on the environment, it is convenient to use an open quantum system description where the full Hamiltonian is partitioned as
\begin{equation}
    H=H_\mathrm{S}+H_\mathrm{B}+H_\mathrm{SB},
\end{equation}
where $H_\mathrm{S}$ describes the system,  $H_\mathrm{B}$ the bath (environment), and  $H_\mathrm{SB}$ the interaction between system and bath.
To evaluate the correlation function (Eq.~\ref{eq:correlation_general}) it is then not necessary to explicitly propagate all individual bath-degrees of freedom, but one can use an effective description of the environment. 
For a weak-system bath coupling and/or Markovian environments the full Liouville space propagator $\eul^{-\ii \mathcal{L} t}$ can be replaced by a propagator $\eul^{-\ii \mathcal{K} t}$ in the system space. 
The operator $\mathcal{K}$ is determined e.g.\ by a Lindblad or Redfield quantum master equation.
However, in many cases of interest the influence of the environment is non-Markovian and the coupling is not necessarily weak.  
Then one has to use more advanced methods, as mentioned in the introduction.

In the following we assume that the environment can be described by a collection of bosonic modes with frequency $\omega_k$ and creation (annihilation) operators $b_k^\dagger$ ($b_k$).
These modes are linearly coupled to a system operator $S$.
\begin{align}
    \label{ham} 
    H_\mathrm{B} &= \sum_k\,\omega_k b_k^{\dagger}b_k
    , 
    \\  H_\mathrm{SB}&=S\otimes B=S\otimes
    \sum_k(g_k b_k + g_k^{*}b_k^{\dagger}).
\end{align}
 The system-bath coupling constants $g_k$ define the  spectral density $J(\omega)~=~\sum_k\abs{g_k}^2\delta (\omega - \omega_k)$, which is related to the bath-correlation function $\alpha(t-s)=\langle B(t) B(s)\rangle $ via
 \begin{equation}
     \alpha(t)=\int_0^\infty  \Big(\coth (\tfrac{\beta \omega}{2})\cos(\omega t) - \ii \sin(\omega t)  \Big)\, J(\omega)\, \mathrm{d}\omega,
 \end{equation}
 where $\beta$ is the inverse temperature. 

\section{Correlation functions from $\mathrm{uni}$TEMPO\label{sec:UniTEMPO}}

For non-Markovian dynamics, multi-time correlation functions can be computed numerically using exact methods such as HEOM \cite{tanimuraNumericallyExactApproach2020}, Hierarchy of pure states (HOPS) \cite{suessHierarchyStochasticPure2014}, chain mappings \cite{Hughes2009Jul, tamascelliEfficientSimulationFiniteTemperature2019}, pseudomodes \cite{Imamog1994, Garraway1997, Roden_2011,  Mascherpa2020May}, quasi adiabatic propagator path integral (QUAPI) \cite{makriTensorPropagatorIterative1995}, small matrix path integral (SMAT-PI) \cite{makriSmallMatrixDisentanglement2020a} or tensor network methods. Most of these approaches realize memory effects via embedding the full non-Markovian dynamics within a larger state space including "auxiliary" memory degrees of freedom. These are constructed in such a way that the combined evolution of auxiliary and system degrees of freedom is strictly time-local, but the reduced system dynamics matches the original non-Markovian model. Approaches based on tensor network influence functionals such as PT-TEMPO, uniTEMPO or the automated compression of environments (ACE) method \cite{Cygorek2021Jan}, can automatically generate such an auxiliary dynamics for structured environments with a user-defined accuracy threshold, where higher accuracy requires larger auxiliary spaces. The uniTEMPO method that we use in this work exploits time-translation invariance of the bath correlation function $\alpha(t-s)$ in order to construct a time-independent auxiliary space, allowing for a direct propagation to arbitrary evolution times \cite{Link_PhysRevLett.132.200403}. 

\begin{figure}[t]
    \centering
    \includegraphics[width=8.6cm]{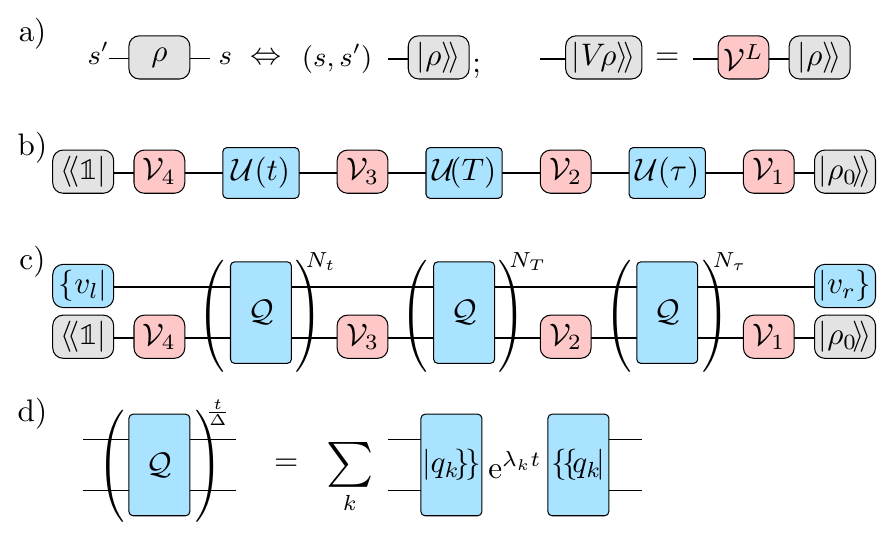}
    \caption{Tensor network diagrams. a) Vectorization of a density matrix. The matrix indices are combined to a single superindex. Operators acting on the density matrix are transformed to superoperators. b) A 4-point correlation for a Markovian system. c) Non-Markovian 4-point correlation expressed in an extended state space via uniTEMPO. d) The spectral decomposition of the non-Markovian propagator used in this work.}
\label{fig:diagrams}
\end{figure}

In the uniTEMPO approach, for a discrete time-step $\tau=\Delta$  (Trotter step), the system propagator $\mathcal{U}(\Delta)$ from Eq.~(\ref{eq:correlation_general}) is replaced by a dissipative propagator $\TOp{Q}$ on an extended state space. It can be represented as a matrix acting on vectors $\Fket{r}\in \mathbb{C}^{d^2\times \chi}$, where $d$ is the dimension of the system Hilbert space and $\chi$ the dimension of the auxiliary space. The dimension of $\TOp{Q}$ is therefore $(d^2\cdot \chi)\times (d^2\cdot \chi) $. Note that the auxiliary space is not a physical state space, i.e.~$\chi$ is in general not a square number and $\TOp{Q}$ is not a quantum channel. For this reason we use the notation $\Fket{\cdot}$ instead of $\Lket{\cdot}$ for state vectors in the extended and auxiliary spaces.

For time-independent system Hamiltonians (as we have), $\TOp{Q}$ is the same for all time steps. Thus, for a given initial vector $\Fket{r_\mathrm{ini}}$,  the state at time $\tau=N_\tau \Delta$ is (cf.~Eq.(\ref{eq:rho^nu(t)}))
\begin{equation}
 \Fket{r(\tau)}= \TOp{Q}^{N_\tau}\Fket{r_\mathrm{ini}}= \TOp{Q}^{\tfrac{\tau}{\Delta}}\Fket{r_\mathrm{ini}}.   
\end{equation}
Together with the short-time propagator $\TOp{Q}$ the uniTEMPO algorithm also provides effective environments vectors $\Tket{\TOp{v}_\mathrm{R}}$ and $\Tbra{\TOp{v}_\mathrm{L}}$ that represent the initial environment and an effective trace over the environment, respectively.
The trace of a time-evolved state can then be expressed as
$\Fbraket{\openone}{r(\tau)}$ where $\Fbra{\openone}=\Lbra{\openone} \otimes\{\TOp{v}_\mathrm{L}|$ represents the trace operation in the extended space. 
For uncorrelated initial conditions the initial state is given by $\Fket{r_\mathrm{ini}}= \Lket{\rho^\mathrm{sys}_0}\otimes \Tket{\TOp{v}_\mathrm{R}}$.

The computation of multi-time correlations in this framework is straightforward. 
The operators $\mathcal{V}_j$ only interact with the system degrees of freedom so that we can trivially extend them to the enlarged space via $\TOp{V}_j=\mathcal{V}_{j}\otimes \openone_{\chi\times\chi}$. 
The multi-time correlation function is then written as
\begin{equation}
\label{eq:R_UniTEMPO}
    R(\tau_n,\dots,\tau_1)
    =
    \Fbra{\openone}\TOp{V}_{n+1}\TOp{Q}^{N_{\tau_{n}}}  \cdots  \mathsf{V}_2  \mathsf{Q}^{N_{\tau_1}} \mathsf{V}_1 \Fket{r_\mathrm{ini}},
\end{equation}
where $\tau_j=N_{\tau_j} \Delta$. 
A diagrammatic representation of this equation is provided in Fig.~\ref{fig:diagrams}c).

To further facilitate the calculations, it is convenient to work in the eigen-decomposition of $\TOp{Q}$, which is given by
\begin{equation}
 \TOp{Q}= \sum_{k=1}^{d^2\cdot\chi} q_k \Fket{q_k}\Fbra{q_k},   
\end{equation}
where the $q_k$ are in general complex numbers and the $\Fket{q_k}$  and  $\Fbra{q_k}$  are the corresponding right and left eigenvectors.
With this, Eq.~(\ref{eq:R_UniTEMPO}) becomes
\begin{align}
\label{eq:R_q}
     &R(\tau_n,\dots,\tau_1)
     \\
    &\phantom{a}=\!
    \sum_{k_n,\dots,k_1}\Fbraket{\openone}{\TOp{V}_{n+1}|q_{k_n}} q_{k_n}^{N_{\tau_{n}}} \cdots    \Fbra{q_{k_2}}\TOp{V}_2\Fket{q_{k_1}}  q_{k_1}^{N_{\tau_1}} \Fbra{q_{k_1}}\TOp{V}_1\Fket{r_\mathrm{ini}}.\nonumber
\end{align}

To connect to the common formulation, where the time-evolution between the interaction has an exponential form, we note that
\begin{equation}
    q_k^{N_t}=q_k^{t/\Delta}\equiv \eul^{\lambda_k t},\qquad \mathrm{with} \quad \lambda_k=\log(q_k)/\Delta
    \label{eq:lambda_k=log}
\end{equation}
where $\lambda_k$ is in general complex.
We express $\lambda_k$ in terms of real frequencies $\omega_k$ and real positive damping rates $\gamma_k$: 
\begin{equation}
    \lambda_k= -\ii \omega_k - \gamma_k, \quad \mathrm{with}\ \gamma_k>0.
\end{equation}
This reformulation is particularly useful when one is interested in the frequency-domain expressions. 
Then a half-sided Fourier transformation of the time interval $\tau$ simply results in the replacement of $q_{k}^{N_{\tau}} $ in 
Eq.~(\ref{eq:R_q}) by $\int_0^\infty \eul^{\pm \ii \omega_{\tau}\tau} \eul^{\lambda_{k} \tau} \mathrm{d}\tau=\frac{-1}{\pm \ii \omega_{\tau}+\lambda_{k}}$.

\section{multi-dimensional spectroscopy\label{sec:2D-spek}}

In coherent multi-dimensional electronic spectroscopy several laser pulses with specific phases, polarizations and directions couple via the dipole operator to the system. 
For pulse $j$ we denote the corresponding interaction operator as $V_j$.
As mentioned in the introduction, these operators act only on the system degrees of freedom (i.e.\ electronic states). It is common to denote the first time interval $\tau_1$ by $\tau$ (coherence time), the second time interval $\tau_2$ by $T$ (waiting time) and the third time interval $\tau_3$ by $t$ (detection time).

To keep the discussion transparent, we work in the semi-impulsive limit, neglecting the temporal width of the laser pulses~\cite{Mu95__}. 
Furthermore, we make the rotating-wave approximation and assume that the laser pulses  are time-ordered~\cite{Mu95__}. 
Then, only the following four multi-time correlation functions contribute to the signal: 
\begin{equation}
\begin{split}
\label{eq:Response_function_2D_spec}
    R_1=&\Lbra{\openone}\mathcal{V}_4^L\,\mathcal{U}(t)\,\mathcal{V}_3^R\, \mathcal{U}(T)\,\mathcal{V}_2^R\, \mathcal{U}(\tau)\, \mathcal{V}_1^L \Lket{\rho_0},
    \\
R_2=&\Lbra{\openone}\mathcal{V}_4^L\,\mathcal{U}(t)\,\mathcal{V}_3^R\, \mathcal{U}(T)\,\mathcal{V}_2^L \,\mathcal{U}(\tau) \,\mathcal{V}_1^R \Lket{\rho_0},
\\
R_3=&\Lbra{\openone}\mathcal{V}_4^L\,\mathcal{U}(t)\,\mathcal{V}_3^L\, \mathcal{U}(T)\,\mathcal{V}_2^R\, \mathcal{U}(\tau) \,\mathcal{V}_1^R \Lket{\rho_0},
\\
R_4=&\Lbra{\openone}\mathcal{V}_L^L\,\mathcal{U}(t)\,\mathcal{V}_3^L \,\mathcal{U}(T)\,\mathcal{V}_2^L \,\mathcal{U}(\tau)\, \mathcal{V}_1^L \Lket{\rho_0},
\end{split}    
\end{equation}
where $\mathcal{V}_j^R\Lket{\rho}\leftrightarrow \openone \rho V_j$ describes the right-interaction and $\mathcal{V}_j^L\Lket{\rho}\leftrightarrow V_j \rho \openone$ the left-interaction of the operators $V_j$ defined in Sec.~\ref{sec:general_def}.

A variety of different signals are possible, depending on the experimental technique (e.g.\ phase-matching / phase-cycling), which theoretically are reflected in different sums of certain Liouville-space pathways. 
In the exemplary calculations shown below we will present the four individual response functions.

Typically 2D-spectra are represented by transforming the first ($\tau$) and last ($t$) time interval into frequency domain, resulting in two-dimensional spectra for each waiting time $T$.
We use the following Fourier transform convention 
\begin{equation}
\label{eq:S_j(R_j)}
    S_j(\omega_t,T,\omega_\tau)= \int_{0}^\infty   \int_{0}^\infty           R_j(t,T,\tau) \eul^{\ii \omega_t t}e^{\ii \sigma_j \omega_\tau \tau} \mathrm{d}t \mathrm{d}\tau,
\end{equation}
where  $\sigma_j\in \pm 1$ is the sign entering the Fourier transformation with respect to the coherence time $\tau$.
Explicitly we use $  \sigma_j=+1$ for $j=1,\,4$ and $\sigma_j=-1$ for $j=2,\,3$.

We also determine the linear spectrum from
\begin{equation}
    L(\omega)= \int_0^\infty \Lbra{\openone}\mathcal{V}^\mathit{L}_2\;\mathcal{U}(\tau)\,\mathcal{V}_1^\mathit{L}\Lket{\rho_0} \eul^{\ii \omega \tau}\, \mathrm{d}\tau.
\end{equation}

\section{Linear- and 2D-spectra from $\mathrm{uni}$TEMPO\label{sec:2D-unitempo}}

\subsection{General formula}
According to Eq.~(\ref{eq:R_q}) the half-sided Fourier transformations give simply (complex) Lorentzian functions for each eigenvalue.
Therefore, we can write for the (complex) linear spectrum 
\begin{equation}
\label{eq:Lin_spek}
L(\omega)
=
\sum_{k}\frac{\Fbraket{1|\TOp{V}_2}{q_k} \Fbra{q_k}\TOp{V}_1\Fket{r_\mathrm{ini}} } {\ii(\omega- \omega_k)-\gamma_k},
\end{equation}
and for the (complex) 2D-spectrum 
\begin{equation}
\label{eq:2d-spectrum-T}
 S_j (\omega_t, T, \omega_\tau) = \sum_{kn} \frac{\Fbra{1}\TOp{V}_4\Fket{q_k}\sum_l(\TOp{V}_3^{kl}q_l^{N_T}  \TOp{V}_2^{ln} )  \Fbra{q_n}\TOp{V}_1\Fket{r_\mathrm{ini}} } {(\ii(\omega_t-\omega_k)-\gamma_k)( \ii(\sigma_j\omega_\tau- \omega_n) -\gamma_n)   },
\end{equation} 
where we have used the abbreviation $\TOp{V}_j^{kl}=\Fbra{q_k}\TOp{V}_j\Fket{q_l}$, and note that $q_l^{N_T} = \eul^{-\ii \omega_l T - \gamma_l T}$.

\subsection{Example calculations \label{ex_cal}} 
We now show some exemplary calculations to demonstrate the applicability of the uniTEMPO method for the calculation of 2D-spectra in various parameter regimes.

We use the same simple model as in Ref.~\cite{deWitt_PhysRevResearch7013209}.
The model consists of a three-level system where only the excited states are coupled to the bosonic bath.
Explicitly we have
\begin{align}
    \label{ham} 
    H_\mathrm{S} =& \big(\epsilon + \lambda\big)\Big( \dyad{1}{1} + \dyad{2}{2}\Big) + \Omega\Big(\dyad{1}{2} + \dyad{2}{1}\Big),\\
    S=&\ket{1}\bra{1} -\ket{2}\bra{2}
\end{align}
The energy of the excited states $\ket{1}$ and $\ket{2}$ is given by the bare energy $\epsilon$ plus the reorganization energy of the bath $\lambda$ (defined below), and electronic coupling $\Omega$. 
We assume that  the laser pulses  couple to the system via the same operator 
\begin{equation}
V_j=V=\ketbra{0}{2}+\ketbra{2}{0}.
\end{equation}
We have ignored coupling strength, field strength and polarizations, since they would show up just as an overall scaling of the signal strength. 

As in Ref.~\cite{deWitt_PhysRevResearch7013209}   we consider $\Omega = 0.2\,\mathrm{ps}^{-1}$ and $\Omega =2.0\,\mathrm{ps}^{-1}$. 
 For the spectral density we use
\begin{equation}
    J(\omega) = 2 \alpha \omega e^{-\frac{\omega}{\omega_c}},
    \label{sd}
\end{equation}
with the cut-off frequency $\omega_c=3.04\,\mathrm{ps}^{-1}$ and the  dimensionless parameter $\alpha$  related to the reorganization energy  by 
\begin{equation}
    \lambda = \int_0 ^\infty \dd\omega \frac{1}{\omega} J(\omega) = 2\alpha\omega_c.
\end{equation}
We set the initial state $\rho_0$ to be a product state of the system ground state $\ket{0}$ and the thermal state of the environment at temperature $1/\beta =13\,\mathrm{ps}^{-1}$. 
For the spectra shown below, we choose $\epsilon$ (which leads just to a global energy shift) as the zero of energy.
I.e.\, we set $\epsilon=0$.

\begin{figure}[htp]
    \centering
    \includegraphics[width=9cm]{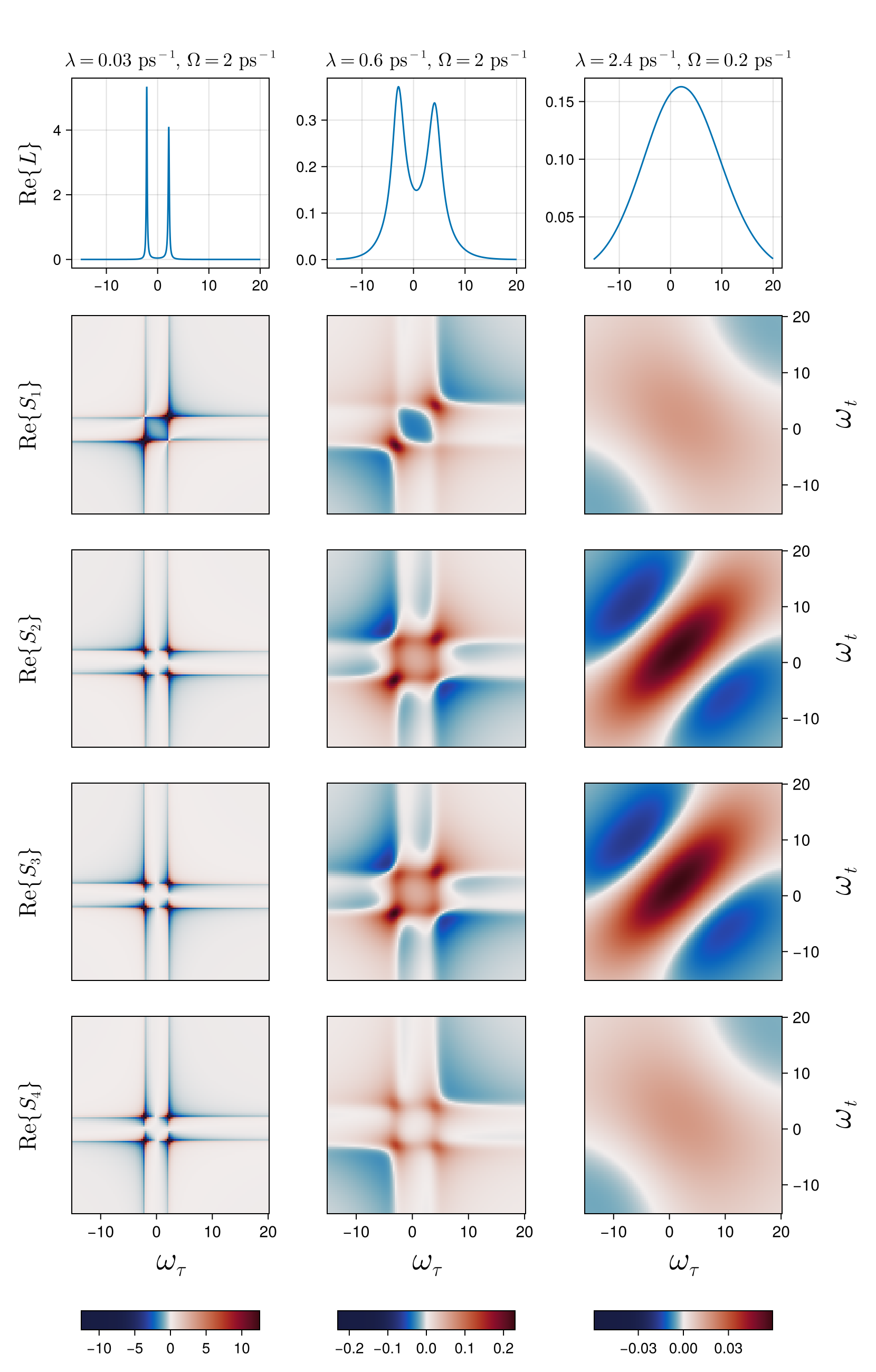}
    \caption{Linear and third order response functions for three different sets of parameters, as indicated on top of each column.
    The upper row shows $\Re{L(\omega)}$ calculated according to Eq.~\,(\ref{eq:Lin_spek}).
    In the four bottom rows 2D-spectra for waiting time $T=0$ are shown.
    These spectra show $\Re{S_j(\omega_t,0,\omega_\tau)}$, as defined in Eq.\,(\ref{eq:S_j(R_j)}), and  are calculated according to Eq.\,(\ref{eq:2d-spectrum-T}). The propagators $\TOp{Q}$ for each column are computed with a time step $\Delta=0.025 \text{ ps}$, yielding auxiliary space dimensions $\chi =21$, $98$, and $301$, respectively.}
    \label{fig:2d_keeling} 
\end{figure}

\begin{figure}[htp]
    \centering
    \includegraphics[width=8.5cm]{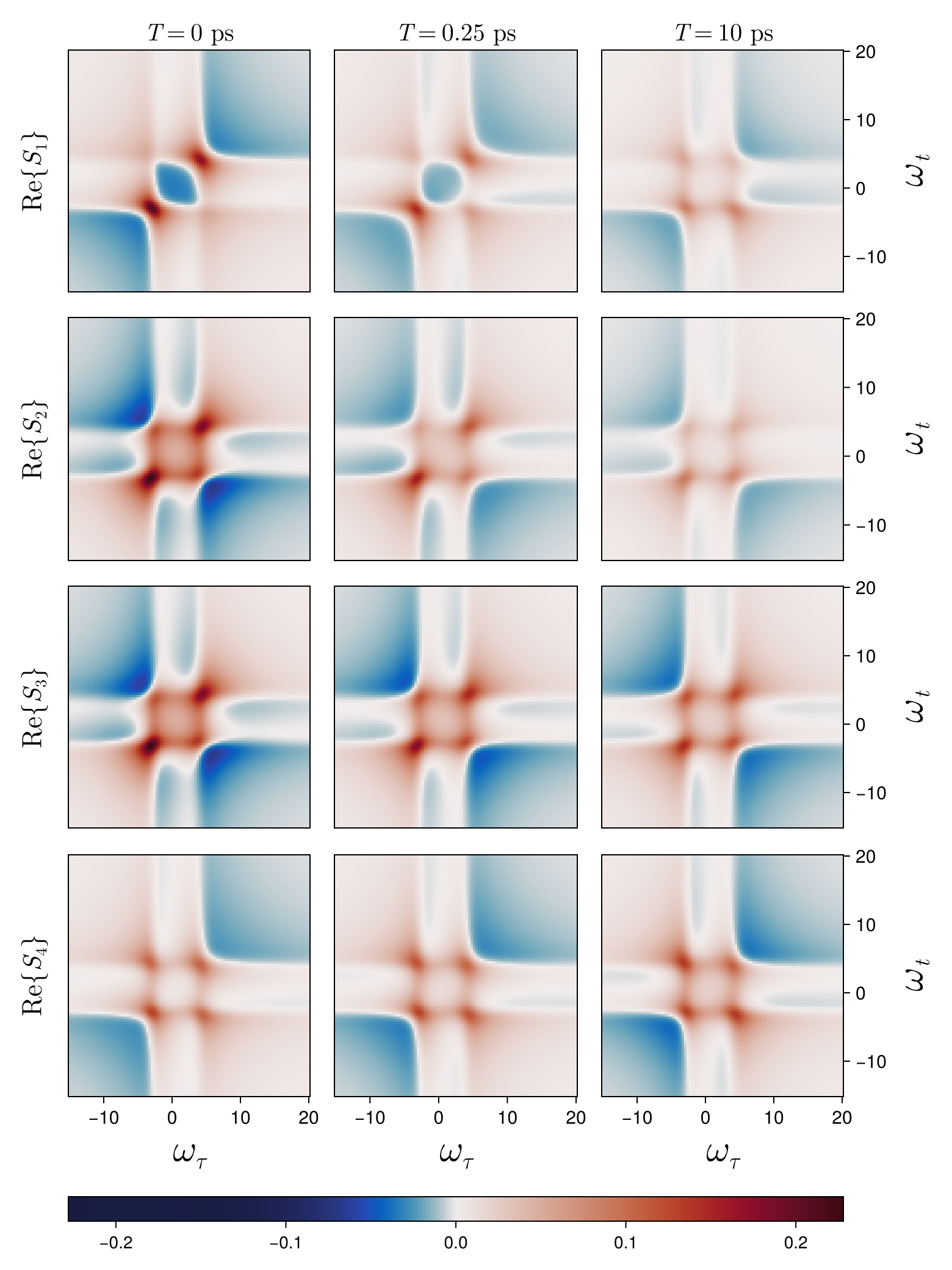}
    \caption{2D-spectra for different waiting times $T$.  
    The parameters are the same as in the middle column of Fig.~\ref{fig:2d_keeling}, i.e.\ $\lambda=0.6\,\mathrm{ps}^{-1}$ and $\Omega=2.0\,\mathrm{ps}^{-1}$. }
    \label{fig:2d_keeling_waiting_time_interm}
\end{figure}

Figure~\ref{fig:2d_keeling} presents spectra for three distinct parameter sets of $\lambda$ and $\Omega$, each representing a different coupling regime as described in Ref.~\cite{deWitt_PhysRevResearch7013209}.
    The left column ($\lambda=0.03\,\mathrm{ps}^{-1}$, $\Omega=2\,\mathrm{ps}^{-1}$) illustrates weak system-bath coupling, the middle column ($\lambda=0.6\,\mathrm{ps}^{-1}$, $\Omega=2\,\mathrm{ps}^{-1}$) intermediate coupling, and the right column ($\lambda=2.4\,\mathrm{ps}^{-1}$, $\Omega=0.2\,\mathrm{ps}^{-1}$) strong coupling.
    The first row displays $\Re{L(\omega)}$, which, up to prefactors, corresponds to the absorption spectra.
    Below each linear spectrum, the corresponding 2D-spectra $\Re{S_j(\omega_t, T=0, \omega_{\tau})}$ are shown, calculated using Eq.~(\ref{eq:2d-spectrum-T}) with a waiting time of $T=0$.
    Notably, Eq.~(\ref{eq:2d-spectrum-T}) allows for the calculation of 2D spectra at arbitrary waiting times $T$ with minimal additional numerical effort. 
Figure~\ref{fig:2d_keeling_waiting_time_interm} extends this analysis by presenting spectra for finite waiting times, focusing on the intermediate coupling case ($\lambda=0.6\,\mathrm{ps}^{-1}$, $\Omega=2.0\,\mathrm{ps}^{-1}$).
    The left column reproduces the $T=0$ spectra from the middle column of Fig.~\ref{fig:2d_keeling}, while the middle and right columns show spectra for waiting times $T=0.25\,\mathrm{ps}$ and $T=10.0\,\mathrm{ps}$, respectively.

\subsection{Some remarks about performance and convergence}

In Appendix \ref{app:unitempo} we show an example for a convergence analysis of linear spectra with respect to the accuracy parameters of uniTEMPO, the bond dimension $\chi$ (i.e.\ the dimension of the auxiliary space) and the Trotter step size $\Delta$.

We find that the Trotter error scales as $O(\Delta^2)$ consistent with the second-order Trotter decomposition. 
For the bond dimension we find that the mean error for linear spectra in the relevant frequency regime follows a scaling $O(\chi^{-r})$ with $r\approx 3$.

The dynamical semi-group structure of the extended dynamics generated via uniTEMPO yields significant advantage in numerical performance over the process tensor time
evolving matrix product operators (PT-TEMPO) algorithm \cite{Jorgensen2019Dec}, where time-translation invariance is lost due to the use of a finite evolution time grid. 
This performance advantage is evident from the formal numerical scaling for computing spectra, as outlined below. 
Note that the bond dimensions required for a given accuracy in PT-TEMPO and uniTEMPO are comparable \cite{Link_PhysRevLett.132.200403}.

In order to compute a spectrum in the frequency domain via uniTEMPO one first diagonalizes the propagator $\TOp{Q}$ with cost $O((\chi d^2)^3)$. This has to be done only once at the beginning of the calculation.

Linear spectra are obtained from the computed eigenvectors and eigenvalues on a frequency grid of size $M$, which adds $O((\chi d^2)M)$, corresponding to a sum over $(\chi d^2)$ eigenvalues (frequencies) in Eq.~\eqref{eq:Lin_spek}.

For 2D-spectra, in order to evaluate Eq.~\eqref{eq:2d-spectrum-T}, one first computes separately the sum over $l$ at $O((\chi d^2)^3)$. 
In the remaining double sum over eigenvalues each summand adds a contribution to the spectrum on the given $M\times M$ frequency grid. The numerical scaling for this last step is thus $O((\chi d^2)^2M^2)$. 
By contrast, the scaling of PT-TEMPO, where real-time evolution is required on a quadratic time-grid, has an a priori scaling $O((\chi d^2)^3 N^2)$, where $N$ is the number of time steps.
Note also that in uniTEMPO only the single tensor $\TOp{Q}$ needs to be stored, so that memory requirements do not grow with propagation time.

Numerical scaling alone often does not reflect the practical performance of a method. We therefore also report computation times for our example simulations, performed on a single core of an Intel i9 (13th Generation) processor. In general, the computation of our example spectra is computationally inexpensive, with execution times below a few minutes and negligible memory demand. Table~\ref{tab:computation-time} reports the computation times in seconds for both constructing the propagator $\TOp{Q}$ and computing the response function $S_1$ (single pathway) of the second row of Figure~\ref{fig:2d_keeling}. 
The calculation of spectra for a finite waiting time $T$ also does not significantly increase the computation time, as the time propagation is efficiently carried out via exponentiation of the eigendecomposition of the propagator $\TOp{Q}$. 

\begin{table}[h!]
\centering
   \caption{Computation times (in seconds) for constructing the propagator $\TOp{Q}$ and computing the response function $S_1$ (single pathway) over a $100\times100$ frequency grid. The different values of the auxiliary space (bond) dimension $\chi$ correspond to the respective columns of the second row of Figure~\ref{fig:2d_keeling}. All timings were measured on a single core of an Intel i9 (13th Generation) processor.}
   
    \renewcommand{\arraystretch}{1.2}
    \begin{tabular}{l|c|c|c|}
        \toprule
        Auxiliary space dimension ${\chi}$ & {\hspace{5pt}$21$\hspace{5pt}} & {\hspace{5pt}$98$\hspace{5pt}} & {\hspace{4pt}$310$\hspace{4pt}} \\
        \midrule
        \hline
        Propagator $\TOp{Q}$       & 0.2 & 1.1 &  6.3 \\
        Response function    & 0.06 & 2.2 & 87 \\
        \bottomrule
    \end{tabular}
    \label{tab:computation-time}
\end{table}

\section{Conclusions\label{sec:conclusions}}

In this study, we showed how to compute multi-time correlation functions for non-Markovian open quantum systems using the uniTEMPO approach.
Calculation of the  corresponding spectra in frequency domain is highly efficient as the necessary Fourier transforms can be performed analytically. 

We demonstrated the efficiency of the approach for a simple system, presenting scaling relations and CPU times.
Our results highlight the computational power of uniTEMPO.
A particularly attractive feature of uniTEMPO is its ability to calculate 2D-spectra for large waiting times $T$ without  additional computational cost.

While our example calculations focused on correlation functions with three time intervals, we emphasize that extending the method to higher-order correlation functions—now increasingly relevant in femtosecond spectroscopy—only moderately increases numerical demands. 
This is because the short-time propagator $\TOp{Q}$ is calculated and diagonalized just once. Each additional order introduces only an extra summation over the dimension of the auxiliary space $\chi$ times the square of the system Hilbert space dimension.

Our current analysis considered a single system operator coupled to the environment. As many relevant systems involve multiple system operators interacting with multiple (potentially correlated) environments, our next step will be to explore this more complex scenario. Beyond multi-dimensional spectroscopy, we expect the uniTEMPO approach to correlation functions to be broadly applicable, including, for example, to quantum transport.

\section*{Acknowledgment}
We would like to thank Jianshu Cao for the many enlightening discussions on open quantum dynamics we have had with him over many years.

AE acknowledges support from the DFG via a Heisenberg fellowship (Grant No EI 872/10-1).

\section*{Data availability}
The data that support the findings of this study are available from the corresponding author upon reasonable request. 

\appendix
\section{The uniTEMPO framework\label{app:unitempo}}

In this Appendix we provide details of the uniTEMPO framework as needed for the results of the main text. 
We follow  \cite{Link_PhysRevLett.132.200403} where the original derivation is given and where further information can be found.

\subsection{Time evolution within uniTEMPO}
We consider the standard open system Hamiltonian
\begin{equation}\label{eq:hamilt}
    H(t)=H_\mathrm{sys}\otimes \openone_\mathrm{B}+S\otimes B(t),
\end{equation}
where $H_\mathrm{sys}$ and $S$ are time-independent hermitian operators in the Hilbert space of the system and $B(t)$ is an operator that describes the collective degrees of freedom of a Gaussian environment
consisting of a continuum of bosonic modes \cite{MaKu11__}. 
This operator is characterized by the so-called bath correlation function
$\alpha(t,s)=\tr [\rho_\mathrm{env}(0)B(t)B(s)]$,
where $\rho_\mathrm{env}(0)$ is a Gaussian environment initial state \footnote{We assume without loss of generality that $\tr\rho_\mathrm{env}(0)B(t)=0$.}.
The Hilbert space time evolution operator from time $s$ to time $t$ is $U_\mathrm{sys}=e^{-\mathrm{i}H_{\mathrm{sys}}(t-s)}$.
As in the main text we use a Liouville-space (density matrix space) notation, where the system time-evolution superoperator is given by $\mathcal{U}_\mathrm{sys}=e^{-\mathrm{i}\mathcal{L}_\mathrm{sys}(t-s)}$ with $\mathcal{L}\bullet=[H_\mathrm{sys},\bullet]$.
We work in an eigenbasis of the coupling operator $S$.
Using this eigenbasis we write basis states in Liouville-space as $\Lket{\mu}\equiv \Lket{\mu_\mathrm{L}\mu_\mathrm{R}}$, where  $\mu_\mathrm{L}$ and $\mu_\mathrm{R}$ label the eigenstates of $S$.
The state $\Lket{\mu_\mathrm{L}\mu_\mathrm{R}}$ corresponds to the Hilbert space operator $\ket{\mu_\mathrm{L}}\bra{\mu_\mathrm{R}}$.
The corresponding elements of the density matrix $\rho$ can thus be labeled by an index that runs from $1$ to $d^2$, where $d$ is the dimension of the system Hilbert space.

In Liouville-space, the matrix-elements in the eigenbasis of $S$ are denoted by
\begin{equation}
    \mathcal{U}_\mathrm{sys}^{\nu\mu}(t,s)=\Lbra{\nu}e^{-\mathrm{i}\mathcal{L}_\mathrm{sys}(t-s)}\Lket{\mu}.
\end{equation}

In the following, time is discretized in multiples of the time-step $\Delta$.
  The time evolution of the system state $\rho(t)$ \textit{under the influence of the environment} can then be expressed in terms of a discrete path integral \cite{feynman63,makriTensorPropagatorIterative1995,Jorgensen2019Dec,Cygorek2021Jan}
\begin{equation} 
\begin{split}\label{eq:time_evo}
&\rho^{\nu_N}(N\!\Delta)=\hspace{-2mm}\sum\limits_{\substack{\mu_1...\mu_N \\ \nu_0...\nu_{N-1}}}\hspace{-2mm}\mathcal{F}_N^{\mu_N\ldots\mu_1}\Big(\prod_{k=1}^N\mathcal{U}_{k}^{\nu_{k}\mu_k\nu_{k-1}}\Big)\rho^{\nu_0}(0),
\end{split}
\end{equation}
Here, $\mathcal{F} $  denotes the influence functional, which is the basic quantity describing the environment.
The tensors $\mathcal{U}_k$ describe purely unitary evolution during time-step $k$  and (for a time-independent Hamiltonian) are given by
\begin{equation}
    \mathcal{U}_{k}^{\lambda\mu\nu}=\mathcal{U}^{\lambda\mu}_\mathrm{sys}(k\Delta,(k-\tfrac{1}{2})\Delta)\,\mathcal{U}^{\mu\nu}_\mathrm{sys}((k-\tfrac{1}{2})\Delta,(k-1)\Delta).
\end{equation}
Since we consider a time-independent system Hamiltonian, the unitary evolution is independent of the time step $\mathcal{U}_k^{\lambda\mu\nu}\equiv \mathcal{U}^{\lambda\mu\nu}$.

As shown in Ref.~\cite{Link_PhysRevLett.132.200403} one can write
\begin{equation}\label{eq:F_aux}
\mathcal{F}_N^{\mu_N...\mu_1}=\vec{\TOp{v}}_\mathrm{L}^T f^{\mu_N}\cdots f^{\mu_2} f^{\mu_1}\vec{\TOp{v}}_\mathrm{R} ,
\end{equation}
where, for given index $\mu$, $f^\mu$ is a square matrix (whose dimension we denote by $\chi\times\chi$) and $\Vec{\TOp{v}}_{\mathrm{L}/\mathrm{R}}$ are vectors realizing finite-time boundary conditions. 
 The tensors $f$ are all identical and independent of $N$.

For Gaussian bosonic baths the influence functional takes the exact form \cite{feynman63,makriTensorPropagatorIterative1995,Jorgensen2019Dec}
\begin{equation}\label{eq:f_stat}
\begin{split}
&\mathcal{F}_N^{\mu_N...\mu_1}=\prod_{i=1}^N\prod_{j=1}^i I_{i\!-\!j}(\mu_i,\mu_j)
\\
&I_{k}(\mu,\nu)= \exp\left(-(S_{\mu_\mathrm{L}}-S_{\mu_\mathrm{R}})(\eta_k S_{\nu_\mathrm{L}}-\eta_k^*S_{\nu_\mathrm{R}})\right),
\end{split}
\end{equation}
where the discretized bath correlation function is given as
\begin{equation}
    \eta_{k}=  \begin{cases}
      \int_{k\Delta}^{(k+1)\Delta}\diff t\int_{0}^{\Delta}\diff s\,\alpha(t-s), & k>0 \\
      \int_{0}^{\Delta}\diff t\int_{0}^{t}\diff s\,\alpha(t-s), & k=0
    \end{cases}
\end{equation}
and $S_n$ denotes the $n$'th eigenvalue of the coupling operator $S$.

Using the MPO representation of the influence functional in the form Eq.~\eqref{eq:F_aux} we define a short-time propagator for a time step $\Delta$.
Its matrix-elements are given by
\begin{equation}
    \TOp{Q}^{\lambda \nu}_{ij}=  \sum_{\mu=1}^{d^2}\mathcal{U}^{\lambda\mu\nu}\,f^\mu_{ij}.
\end{equation}
Using the matrix $\TOp{Q}$ the time evolution Eq.~\eqref{eq:time_evo} of the density matrix of the system can be written as
\begin{equation}
\label{eq:rho^nu(t)}
    \rho^{\nu}(t)=\sum_{\lambda=1}^{d^2}\sum_{i,j=1}^\chi\,\big[\TOp{G}(t)\big]_{ij}^{\lambda\nu}\,\rho^\lambda(0)\TOp{v}_\mathrm{L}^i\TOp{v}_\mathrm{R}^j,
\end{equation}
where 
\begin{equation}
\TOp{G}(t)=\TOp{Q}^{\tfrac{t}{\Delta}}=\TOp{Q}^N,
\end{equation}
is the $N$-th power of the matrix $\TOp{Q}$ and can be explicitly written as
\begin{equation}
   \big[\TOp{Q}^{N}\big]_{ij}^{\lambda\nu}
   = \sum_{j_1\ldots j_{N-1}=1}^\chi\sum_{\nu_1\ldots\nu_{N-1}=1}^{d^2}
   \TOp{Q}^{\lambda,\nu_{N-1}}_{i,j_{N-1}}\cdots \TOp{Q}^{\nu_2,\nu_1}_{j_2,j_1}
   \TOp{Q}^{\nu_1,\nu}_{j_1,j}.
\end{equation}

\subsection{Eigendecomposition of $\TOp{Q}$}

To calculate the spectroscopic signals it is convenient to use the eigendecomposition of $\TOp{Q}$ in order to evaluate the matrix power $\TOp{Q}^{t/\Delta}$. 
Since $\TOp{Q}$ is non-Hermitian its spectral decomposition takes the form
\begin{equation}
    [\TOp{Q}]_{i,j}^{\lambda,\nu}=\sum_{k=1}^{\chi d^2} q_k [l_k]^\lambda_i [r_k]^\nu_j,
\end{equation}
with complex eigenvalues $q_k\in \mathbb{C}$ and mutually orthogonal left and right eigenvectors $l_m^\dagger r_n=\delta_{mn}$. The corresponding coefficients are obtained from $[r_k]^\nu_j= \Lbra{\nu}\otimes\Tbra{j}\Fket{q_k}$, and similarly for $[l_k]^\lambda_i$.

This way we can decompose the system evolution as
\begin{equation}\label{eq:spectral_decomp}
\begin{split}
        \rho_\mathrm{sys}^{\nu}(t)=&\sum_{k=1}^{\chi d^2} q_k^{t/\Delta}\rho_k^\nu=\sum_{k=1}^{\chi d^2} \eul^{\lambda_k t}\rho_k^\nu,\,\qquad 
        \\
        \rho_k^\nu=&\sum_{\lambda=1}^{d^2}\sum_{i,j=1}^\chi\,[l_k]^{\nu}_j[r_k]^{\lambda}_i \,\rho_\mathrm{sys}^\lambda(0)\TOp{v}_\mathrm{L}^i \TOp{v}_\mathrm{R}^j.
        \end{split}
\end{equation}
We have defined the rates corresponding to each eigenvalue as $\lambda_k = \log(q_k)/\Delta$. Since the density operator is finite for any time we must have $\mathrm{Re}(\lambda_k)\leq 0$. 

\section{Convergence of multitime correlations}

The convergence of the uniTEMPO algorithm is governed by two independent parameters: the Trotter time step $\Delta$ and the bond dimension $\chi$ . We consider here as an example the linear spectrum calculated via Eq.~\eqref{eq:Lin_spek} in the intermediate coupling regime (see the middle column of Fig.\,\ref{fig:2d_keeling}). 
Since we do not have an analytical reference solution, we investigate numerical convergence by quantifying the difference of spectra computed with varying parameters. 
In particular, we quantify the difference of two spectra $L$ and $\tilde L$ calculated with parameters $\Delta$, $\chi$ and $\tilde \Delta$, $\tilde \chi$ via $\delta\epsilon = \frac{1}{n_\omega} \sum_\omega |\tilde L(\omega) - L(\omega)|$ where $n_\omega$ is the number of frequency points.

\begin{figure}[tbp]
    \centering
    \includegraphics{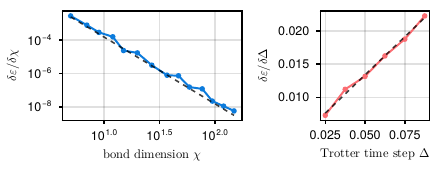}
    \caption{Convergence of the uniTEMPO calculation for the linear spectrum shown in Fig.~\ref{fig:2d_keeling} (intermediate coupling, center panel), for a frequency interval $\omega \in [-15,15]$ with $n_\omega = 5000$ points. The black dashed line indicates $O(\chi^{-(r+1)})$ scaling with $r =3$. Left panel: difference of the linear spectrum with respect to the next-larger bond dimension. The Trotter time step is held fixed at $\Delta = 0.025$. Right panel: difference of the linear spectrum with respect to the next-larger time step at bond dimension $\chi \approx 170$. The black dashed line indicates linear scaling.}
    \label{fig:scaling}
\end{figure}

In the left panel of Fig.~\ref{fig:scaling} the convergence with respect to the bond dimension is displayed. For fixed time step $\Delta = 0.025$ we calculate the difference with respect to the next-larger bond dimension $\tilde\chi$ divided by the bond dimension difference $\delta\chi = |\tilde \chi - \chi|$. This derivative-like quantity allows us to extract the error scaling $O(\chi^{-r})$ without knowledge of the exact solution via 
\begin{equation}
    \frac{\delta\epsilon}{\delta\chi} \sim O\left( 1/\chi^{r+1} \right).
\end{equation}
We find rapid convergence with respect to the bond dimension. For the example in Fig.~\ref{fig:scaling} we extract the exponent $r\approx 3$. 

In the right panel of Fig.~\ref{fig:scaling} the convergence with respect to the Trotter time step $\Delta$ is displayed. We keep the bond dimension roughly constant at around $\chi\approx170$ 
and vary the Trotter time step. We again compute the difference of the spectrum with respect to the next-larger time step $\tilde\Delta$ and divide by the time step difference $\delta \Delta = |\tilde\Delta-\Delta|$. From the second order Trotter expansion we expect an error of $O(\Delta^2)$ for the spectrum, leading to
\begin{equation}
    \frac{\delta\epsilon}{\delta\Delta} \sim O(\Delta),
\end{equation}
as confirmed in Fig.~\ref{fig:scaling}.

\FloatBarrier

\bibliographystyle{journal_v5}
\bibliography{References}

\end{document}